\documentclass[11pt]{article}
\pdfoutput=1
\usepackage{jheppub}

\newcommand{\mathe}{\mathrm{e}}
\usepackage{float}
\usepackage{pgfplots}
\usepackage{tikz}
\usetikzlibrary{arrows.meta}
\pgfplotsset{compat=1.14}

\begin{document}

\preprint{UUITP-49/19}

\title{\boldmath From Exact Results to Gauge Dynamics on $\mathbb{R}^3\times S^1$}

\author{Arash Arabi Ardehali,}
\author{Luca Cassia,}
\author{and Yongchao L\"u}

\affiliation{Department of Physics and Astronomy, Uppsala University,\\
Box 516, SE-751 20 Uppsala, Sweden}

\emailAdd{ardehali@physics.uu.se}\emailAdd{luca.cassia@physics.uu.se}\emailAdd{yongchao.lu@physics.uu.se}

\abstract{We revisit the vacuum structure of the $\mathcal{N}=1$
Intriligator-Seiberg-Shenker model on $\mathbb{R}^3\times S^1$.
Guided by the Cardy-like asymptotics of its Romelsberger index, and
building on earlier semi-classical results by Poppitz and \"{U}nsal,
we argue that previously overlooked non-perturbative effects
generate a Higgs-type potential on the classical Coulomb branch of
the low-energy effective 3d $\mathcal{N}=2$ theory. In particular,
on part of the Coulomb branch we encounter the first instance of a
dynamically-generated quintic monopole superpotential.}

\maketitle \flushbottom

\section{Introduction}

\noindent Some twenty-five years ago, Intriligator, Seiberg, and
Shenker (ISS) \cite{Intriligator:1994rx} published a study of a
fascinating $\mathcal{N}=1$ model which has continued to display
surprising features ever since. The model is an SU(2) gauge theory,
which besides the gauge multiplet only has a single chiral multiplet
$Q_{I=3/2}$ transforming in the spin-$3/2$ representation of the
gauge group. It is asymptotically free (with the one-loop beta-function coefficient $b=6-5=1$), and has a
U(1)$_R$ symmetry under which $Q_{I=3/2}$ has charge $3/5$.

Using holomorphy arguments it was shown that the model, placed on
$\mathbb{R}^{4}$, admits a moduli space of SUSY vacua parametrized
by the basic gauge-singlet superfield $u=Q_{I=3/2}^4$ (with a
totally-symmetric contraction of the gauge indices). As for the
most interesting point $u=0$ on the moduli space, there used to be two main proposals for its IR
behavior: confinement versus non-trivial superconformal fixed-point.
The original study \cite{Intriligator:1994rx} deemed the confinement
scenario more likely because of the remarkable $\mathrm{Tr}R$ and
$\mathrm{Tr}R^3$ 't~Hooft anomaly matchings between the UV constituents and
$u$ as the IR superfield. Under the confinement hypothesis, it was
shown that dynamical SUSY breaking can be realized by adding a
tree-level superpotential.

However, the confinement proposal was subsequently questioned
\cite{Brodie:1998vv,Intriligator:2005if}, and after several years of
suspense eventually ruled out by Vartanov in \cite{Vartanov:2010xj}.
Vartanov's work is an outstanding example of use of exact results to
settle questions in $\mathcal{N}=1$ gauge dynamics. The exact
observable employed in \cite{Vartanov:2010xj} was the Romelsberger
index \cite{Romelsberger:2005eg}, which up to a Casimir-energy
factor can be identified with the supersymmetric partition function
on $S^{3}\times S^{1}$ \cite{Assel:2014paa}. The index is known to be
RG-invariant. Therefore under the confinement scenario of
\cite{Intriligator:1994rx} it should have matched between the IR
superfield $u$ and the UV fields of the ISS model; explicit
computation found otherwise \cite{Vartanov:2010xj}.

Closely related is the dynamics of the ISS model on
$\mathbb{R}^{3}\times S^{1},$ with \emph{periodic} spin structure
on the circle. It has been studied by Poppitz and \"{U}nsal
\cite{Poppitz:2009kz} with the motivation to approach the
$\mathbb{R}^{4}$ behavior via an adiabatic argument by varying
the radius $r_{S^{1}}$ of the circle. Their results include a
description of the vacua on $\mathbb{R}^{3}\times S^{1}.$ In particular, it was suggested in \cite{Poppitz:2009kz} that the moduli space of vacua on
$\mathbb{R}^{3}\times S^{1}$ contains a new flat direction, namely a
compact Coulomb branch (for the low-energy effective 3d
$\mathcal{N}=2$ theory living on $\mathbb{R}^{3}$) parametrized by
a periodic holonomy variable related to the component
of the 4d gauge field along the circle.

In the present paper we employ the Romelsberger index to clarify the
low-energy dynamics of the ISS model on $\mathbb{R}^{3}\times
S^{1}.$ Specifically, our study is motivated by the finding in \cite{Ardehali:2015bla}
that the asymptotic behavior of the Romelsberger index of the model,
as $r_{S^{3}}\to\infty$, exhibits a non-trivial potential for the
holonomy variable (see Figure~\ref{fig:ISS}). In light of this result, and building on earlier
semi-classical developments due to Poppitz and \"{U}nsal, we
re-examine the arguments and conclusions of \cite{Poppitz:2009kz} regarding the compactified ISS model.

Below, by the ``low-energy'' regime on $\mathbb{R}^{3}\times S^{1}$
we mean the regime where energies are much smaller than $1/r_{S^1}$.
Whenever performing semi-classical analysis, we moreover assume that
$r_{S^1}\Lambda_{\mathrm{ISS}}\ll 1$, with $\Lambda_{\mathrm{ISS}}$
the dynamical scale of the model. In such low-energy,
semi-classical regimes, a Kaluza-Klein (KK) reduction on the circle
gives a finite number of fields that are massless on
$\mathbb{R}^{3}$ and constitute the dynamical degrees of freedom,
together with infinite towers of massive KK modes that ought to be
integrated out.

Our main results are as follows. Firstly, at the classical level, we
carefully identify the fields that are massless in the
three-dimensional sense. We find that the mixture of KK charge and
gauge charge leads to extra massless fields on the classical Coulomb
branch. This leads to a new picture of the classical moduli space (see Figure~\ref{fig:pertVac}).
Secondly, at the semi-classical level, we find a monopole-induced
potential on the classical Coulomb branch (see (\ref{eq:semiClassicW})) which has the same qualitative profile
as the holonomy potential arising from the $r_{S^3}\to\infty$ limit
of the Romelsberger index! This monopole-induced potential has the
effect of lifting the parts of the classical Coulomb branch that are
away from the loci supporting charged massless 3d fields.

We emphasize that our use of the index in addressing gauge dynamics
on $\mathbb{R}^{3}\times S^{1}$ is not as direct as Vartanov's use
of it in \cite{Vartanov:2010xj}. Rather, we utilize (the asymptotics
of) the index \emph{as a guide} for a subsequent careful
semi-classical analysis, and it is the latter that yields our main
results. While it is intuitively expected that the
$r_{S^3}\to\infty$ limit of the index might encode semi-classical
dynamics on $\mathbb{R}^3\times S^1$, the details of the connection are somewhat mysterious at the moment; in particular, the asymptotics of the index yields a potential for the holonomy which is piecewise \emph{linear}, whereas the semi-classical (multi-)monopole potential on $\mathbb{R}^3\times S^1$ is piecewise \emph{exponential}, and the relation between the two is far from clear. We will comment more on this point in section~\ref{sec:discussion}.\\

The rest of this paper is organized as follows. In
section~\ref{sec:Rains} we explain how the \emph{leading} asymptotics of the
Romelsberger index might be used as a guide for studying
semi-classical gauge dynamics on $\mathbb{R}^{3}\times S^{1}.$ In
section~\ref{sec:semiClassics} we perform a careful semi-classical
analysis of the low-energy dynamics of the compactified ISS model.
In section~\ref{sec:3dIR} we make some preliminary remarks on the
IR phase of the low-energy effective 3d $\mathcal{N}=2$ theory. In
section~\ref{sec:discussion} we put our results in a wider context
and also comment on a few related open problems.
Appendix~\ref{appendix:FI} discusses the physical content of the
\emph{subleading} asymptotics of the index, while
appendix~\ref{appendix:Shaghoulian} discusses an alternative
perspective on the piecewise-linear potential arising in the leading
asymptotics.

\section{The index perspective}\label{sec:Rains}

\noindent In this section we present intuitive arguments (as in
section~5.4 of \cite{Ardehali:2015bla}) which will guide our
semi-classical analysis in the next section.

We approach $\mathbb{R}^3\times S^1$ by taking the large-$r_{S^3}$ limit of $S^3\times S^1$. The
partition function on the latter space is called the Romelsberger
index \cite{Romelsberger:2005eg}, and can be computed exactly for
any $\mathcal{N}=1$ gauge theory with a U($1$)$_R$
symmetry.\footnote{To avoid subtleties regarding regularization of
the partition function via analytic continuation, we restrict
attention to theories whose chiral multiplets $\chi$ all have
U($1$)$_R$ charges $r_\chi$ in the range $0<r_\chi<2$. Otherwise curvature
couplings induce tachyonic modes (or lack of such couplings when $r_{\chi}=0$ or 2 leads to non-compact
Higgs branches) in the $S^3\times S^1$ path integral, amounting to divergent
results.} Focusing on non-chiral theories with semi-simple gauge
group for simplicity, the asymptotics of the index as the radius of
the $S^3$ is sent to infinity takes the form
\cite{Ardehali:2015bla,Rains:2009,Ardehali:2016kza}
\begin{equation}
\mathcal{I}(\beta)\approx
\mathe^{-\mathcal{E}^{\mathrm{DK}}_0(\beta)}\int_{\mathfrak{h}_{cl}}
\mathe^{-\frac{4\pi^2}{\beta}L_h(\boldsymbol{x})}\, \mathrm{d}^{r_{G}}{x}
\qquad (\text{as
}\beta\to0),\label{eq:indexAsy}
\end{equation}
where $\beta=2\pi r_{S^1}/r_{S^3}$, and
$\mathcal{E}^{\mathrm{DK}}_0(\beta)=\frac{\pi^2}{3\beta}\mathrm{Tr}R$
is the piece studied by Di~Pietro and Komargodski in \cite{DiPietro:2014bca}. The symbol
$\boldsymbol{x}$ stands for the collection $x_1,\dots,x_{r_G}$
parametrizing the unit hypercube in the Cartan subalgebra of the
gauge group, whose rank is denoted $r_G$; the unit hypercube is
denoted by $\mathfrak{h}_{cl}$. (Note that the range of the integration variables $x_j$ is kept fixed, e.g. from $-1/2$ to $1/2$, as $\beta\to0$.) The exponential function
$z_j=\mathe^{2\pi i x_j}$ maps $\mathfrak{h}_{cl}$ to the moduli space of
the eigenvalues of the holonomy matrix $P\exp(i\oint_{S^1}A_4)$,
with $A_4$ the component of the gauge field $A$ along the $S^1$. The
integral in (\ref{eq:indexAsy}) is thus over the ``classical''
moduli space of the holonomies around the circle; hence the
subscript $cl$ in $\mathfrak{h}_{cl}$. When $S^3$ decompactifies to
$\mathbb{R}^3$, this moduli space becomes a middle-dimensional
section of the classical Coulomb branch of the low-energy effective
3d $\mathcal{N}=2$ theory living on $\mathbb{R}^3$; the rest of the
classical Coulomb branch is parametrized by the dual photons (see
e.g. \cite{Aharony:2013dha}), which the index does not see.

Two questions might arise at this stage. First, why is the $\beta\to0$ limit thought of as decompactifying the $S^3$, rather than  shrinking the $S^1$? Second, what are the asymptotic boundary conditions that our approach via decompactification of the $S^3$ imposes on various fields at the infinity of the limiting $R^3$? The answer to the first question is that both perspectives are valid: in the body of the paper we are mainly interested in the ``crossed-channel'' perspective where $\beta\to0$ is thought of as $r_{S^3}\to\infty$, while in the two appendices and parts of section~\ref{sec:discussion} the ``direct-channel'' perspective, where $\beta\to0$ is thought of as $r_{S^1}\to0$,  will also be discussed. A thorough answer to the second question is beyond the scope of the heuristic arguments of the present section; we only note that to have a correspondence with the index the component of the gauge field along the circle must approach a constant matrix, whose holonomy eigenvalues would be in correspondence with the $z_j$ above.

Continuing our study of the index we observe that the integral in (\ref{eq:indexAsy}) localizes in the ``Cardy-like
limit'' $\beta\to0$ to the minima of the \emph{Rains function}
\begin{equation}
\begin{split}
L_h(\boldsymbol{x}):= \frac{1}{2}\sum_{\chi}(1-r_\chi)\sum_{\rho^{\chi}
\in\Delta_\chi}\vartheta(\rho^{\chi}\cdot
\boldsymbol{x})-\sum_{\alpha_+}\vartheta(\alpha_+\cdot
\boldsymbol{x}),\label{eq:LhDef}
\end{split}
\end{equation}
which, as in \cite{Ardehali:2015bla}, we would like to intuitively interpret as a
quantum effective potential on the classical Coulomb branch of the
low-energy effective 3d $\mathcal{N}=2$ theory on $\mathbb{R}^3$. In
Eq.~(\ref{eq:LhDef}) the set of weights of the gauge-group
representation of the chiral multiplet $\chi$ is denoted by
$\Delta_\chi$, and the positive roots of the gauge group by
$\alpha_+$. The function $\vartheta(x)$ in (\ref{eq:LhDef}) is
defined following Rains as \cite{Rains:2009}
\begin{equation}
\vartheta(x):=\{x\}(1-\{x\}),
\end{equation}
with $\{x\}:=x-\lfloor x\rfloor$ the fractional-part function, defined via the more familiar floor function $\lfloor x\rfloor:=\mathrm{max}\{m\in\mathbb{Z}\ |\ m\leq x\}$. Note that while $\vartheta(x)$ is
piecewise quadratic, the Rains function $L_h(\mathbf{x})$ is
piecewise linear thanks to the U($1$)$_R$ ABJ anomaly cancelation.

With the above reasoning we are led to expect that the minima of
the Rains function encode the vacua of the gauge theory on
$\mathbb{R}^3\times S^1$. Of course any non-trivial vacuum structure
due to the dual photons would be invisible in this approach, as the
index is blind to them. Moreover, any non-trivial Higgs branch of the
3d $\mathcal{N}=2$ theory would also escape this approach, because
for any finite $r_{S^3}$ the
curvature couplings lift them.

Let us now illustrate this connection between the Rains function and the $\mathbb{R}^3\times S^1$ vacua in a few concrete examples. A particularly simple example is that of SU($N$) SQCD, with enough matter multiplets---namely $N_f>N$---so that it has a well-defined U(1)$_R$ symmetry and hence a well-defined index. It is a result essentially due to Rains that in this case a certain generalized triangle inequality (proven in \cite{Rains:2009}) involving $\vartheta(\cdot)$ establishes that the Rains function (\ref{eq:LhDef}) is minimized when all $x_j$ are zero (c.f. section~3.1.1 of \cite{Ardehali:2015bla}). This suggests that the theory on $\mathbb{R}^3\times S^1$ has its classical Coulomb branch lifted by quantum corrections, and only a zero-dimensional space of vacua survives at the quantum level. This is in perfect agreement with the semi-classical analysis of the model as discussed in \cite{Aharony:2013dha}. Similar remarks apply to USp($2N$) SQCD, which hence furnishes another simple example.

A more nontrivial illustration is provided by SO($N$) SQCD. For concreteness we focus on the theory with SO($2N+1$) gauge group. (Analogous comments apply to the theory with SO($2N$) gauge group.) Again we assume enough matter multiplets---namely $N_f>2N-1$---so that there is a well-defined U(1)$_R$ and an index. Then, a rather non-trivial identity (presented in Eq.~(3.51) of \cite{Ardehali:2015bla}) involving $\vartheta(\cdot)$ establishes that the Rains function is minimized when one---and only one---of the $x_j$ is non-zero and the rest are zero (see section~3.2.1 of \cite{Ardehali:2015bla}). This one-dimensional space of quantum vacua is again in complete accord with the semi-classical analysis of the $\mathbb{R}^3\times S^1$ dynamics of the model as discussed in \cite{Aharony:2013dha}.

For many theories the vacua on $\mathbb{R}^3\times S^1$ have not yet been studied but the structure of the locus of minima of the Rains function is understood. In such cases the index perspective leads to conjectures regarding the structure of vacua on $\mathbb{R}^3\times S^1$. In section~5.4 of \cite{Ardehali:2015bla} such conjectures were put forward for all SU($N$) $ADE$ SQCD \cite{Intriligator:2003mi} and the
Pouliot theories \cite{Pouliot:1995zc}---in the appropriate range of their parameters such that all their R-charges are in
$]0, 2[$---as well as for the $\mathbb{Z}_3$ orbifold of the SU($N$) $\mathcal{N}=4$ theory. (Some cases with enhanced supersymmetry were also discussed in \cite{Ardehali:2015bla} on which we comment in section~\ref{sec:discussion} where possible obstructions to the heuristic arguments of the present section are also outlined.)

\begin{figure}[t]
\centering
    \includegraphics[scale=1]{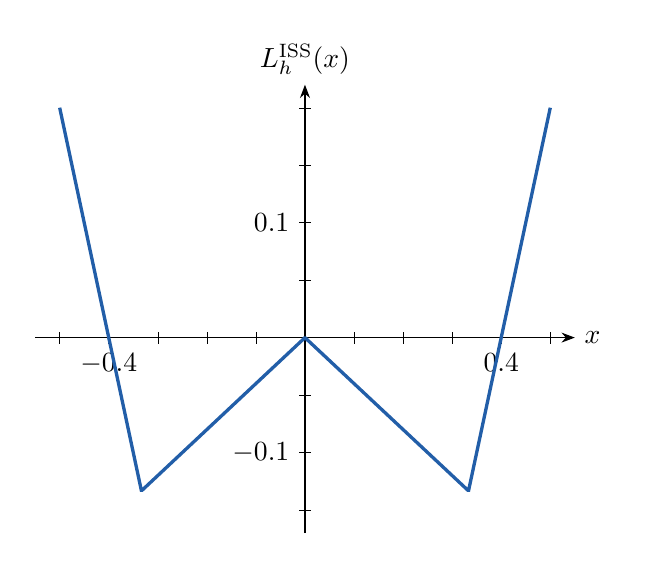}
\caption{The Rains function of the SU(2) ISS model. Its minima lie at $x=\pm1/3$.
\label{fig:ISS}}
\end{figure}

The case of main interest in the present paper is the ISS model, to which we now turn. In terms of $x:=x_1$, the Rains function of the model reads
\begin{equation}
\begin{split}
L^{\mathrm{ISS}}_h(x)=\frac{2}{5}\vartheta(x)
+\frac{2}{5}\vartheta(3x)-\vartheta(2x),
\end{split}\label{eq:LhISS}
\end{equation}
and looks like a Mexican hat as in Figure~\ref{fig:ISS}.

Following the intuitive line of argument advocated above,
Figure~\ref{fig:ISS} would suggest that the \emph{quantum} vacuum of the ISS model on $\mathbb{R}^3\times S^1$ lies at
$x=1/3$ (which is Weyl-equivalent to $x=-1/3$). As
pointed out in \cite{Ardehali:2015bla}, this appears to be in conflict
with an earlier semi-classical analysis of the compactified ISS
model by Poppitz and \"{U}nsal \cite{Poppitz:2009kz}. We now proceed
to argue that the analysis of Poppitz and \"{U}nsal, once extended
slightly further, actually reveals non-perturbative effects
generating a Higgs-type potential qualitatively compatible with
Figure~\ref{fig:ISS}.

\section{Revisiting the semi-classical
analysis}\label{sec:semiClassics}

\noindent We begin our semi-classical analysis with a brief
discussion of the \emph{perturbative} vacuum moduli space of the
compactified ISS model. We assume
$1/r_{S^1}\gg\Lambda_{\mathrm{ISS}}$, so that a semi-classical
analysis is reliable in appropriate regions of the classical moduli
space.

At energies well below the compactification scale $1/r_{S^1}$, semi-classically we have an effective 3d $\mathcal{N}=2$ theory. The scalar in the 3d
$\mathcal{N}=2$ vector multiplet comes from the holonomy of the 4d gauge field as follows. Since the gauge group is SU(2), the two eigenvalues of $P\exp(i\oint_{S^1}A_4)$ are of the form $z^{\pm1}$, and picking one of them as $z$ we can define a periodic scalar $x$ through $z=e^{2\pi i x}$; of course there is a \emph{Weyl redundancy} in picking one of the two eigenvalues as $z$, and throughout the rest of this paper
we fix this redundancy in our scalar by taking $x\in[0,1/2]$.
There is no tree-level potential for this scalar, so it parametrizes
what is called the \emph{classical Coulomb branch} of the 3d
$\mathcal{N}=2$ theory. Actually, for any $x$ strictly inside the
$[0,1/2]$ interval, the super-Higgs mechanism breaks the SU($2$)
gauge group down to U($1$), and the 3d photon can be Hodge-dualized
to a compact scalar $a$, which also does not have a tree-level
potential, and hence combines with $x$ to parametrize the full
classical Coulomb branch.

The Coulomb-branch parameter $x$ also determines the \emph{real
mass} of the 3d descendants of the 4d fields that are charged under
the Cartan of the gauge group. The chiral multiplet $Q_{I=3/2}$
contains fields with charges $-3,-1,+1,+3$ under the Cartan of the
SU($2$). To find the real masses of their 3d descendants we first
note that the $m$th KK mode of a 4d field with charge $n$ under the
Cartan has real mass $(nx+m)/r_{S^1}$; picking the $m\in\mathbb{Z}$
that minimizes the absolute value of this real mass, we get the real
mass of the ``lightest'' among all the KK modes---which we refer to
as \emph{the} 3d descendant.

At $x=0$ all fields yield massless 3d descendants with KK charge $m=0$ of course.
Since the descendants $q_{I=3/2}$ of the chiral multiplet have no
tree-level potential at $x=0$, we have a \emph{classical Higgs
branch} there parametrized by the gauge-invariant combination $u\sim
q_{I=3/2}^4$.

More interestingly, we see that the $x=1/3$ point also supports
massless 3d fields; these are the descendants of the fields with charges
$+3,-3$ inside the chiral multiplet, and we denote them by
$q_{+3},q_{-3}$. (Note that $q_{\pm3}$ have KK charges $m=\mp1.$)
These 3d fields also lack a tree-level potential, and thus yield a
classical Higgs branch at $x=1/3$, which can be parametrized by the
gauge-invariant combination $M\sim q_{+3}q_{-3}$.

Finally, at $x=1/2$ only the 4d vector multiplet yields massless
descendants. The descendants of the 4d gauge multiplet (which have
gauge charges $\pm2$ and KK charges $\mp1$) restore the SU($2$)
gauge group at this point, so we have a 3d pure $\mathcal{N}=2$ SYM
sitting there.

Note that both at $x=0$ and at $x=1/2$ the enhanced SU($2$) is
recovered. At these points the dual photon $a$ ceases to make
sense. In fact even in proximity of these points---in
neighborhoods that we suspect to be more precisely of size $\mathcal{O}(r_{S^1}\Lambda_{\mathrm{ISS}})$ and $\mathcal{O}(r_{S^1}\Lambda_{\mathrm{SYM}})=\mathcal{O}((r_{S^1}\Lambda_{\mathrm{ISS}})^{1/6})$  respectively---we expect our semi-classical
techniques to lose reliability.\footnote{The KK reduction at scale $1/r_{S^1}$ ought to be valid irrespective of $x$ as long as $r_{S^1}\Lambda_{\mathrm{ISS}}\ll1$; it is the semi-classical Higgs picture for $x\neq0,1/2$ that might lose reliability in proximity of the $x=0,1/2$ points. Our emphasis on these subtleties is motivated by the observation that the Rains function also receives correction near $x=0,1/2$; see footnote~3 of \cite{Ardehali:2015bla}. The ``perils of compactification'' discussed in \cite{Marino:1998eg} seem related to these subtleties as well.}

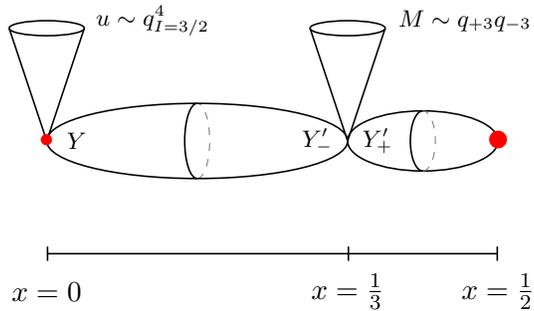
\begin{figure}[t]
\centering
\begin{tikzpicture}
\draw[-,semithick] (0,1.5) -- (-0.5,3); \draw[-,semithick] (0,1.5)
-- (0.5,3); \draw (1.41,3.1) node {\footnotesize$u\sim
q_{I=3/2}^4$}; \draw (.4,1.5) node {\footnotesize$Y$};
\draw[semithick] (-0.5,3) arc (180:0:0.5 and 0.1); \draw[semithick]
(-0.5,3) arc (-180:0:0.5 and 0.1);

\draw[semithick] (2,1) arc (270:90:0.166 and 0.5);
\draw[dashed,color=gray] (2,1) arc (-90:90:0.166 and 0.5);

\draw[semithick] (0,1.5) arc (180:0:2 and 0.5); \draw[semithick]
(0,1.5) arc (-180:0:2 and 0.5);

\draw (3.6,1.5) node {\footnotesize$Y'_-$}; \draw (4.4,1.5) node
{\footnotesize$Y'_+$}; \draw[-,semithick] (4,1.5) -- (3.5,3);
\draw[-,semithick] (4,1.5) -- (4.5,3); \draw (5.56,3.1) node
{\footnotesize$M\sim q_{+3}q_{-3}$}; \draw[semithick] (3.5,3) arc
(180:0:0.5 and 0.1); \draw[semithick] (3.5,3) arc (-180:0:0.5 and
0.1);

\draw[semithick] (4,1.5) arc (180:0:1 and 0.4); \draw[semithick]
(4,1.5) arc (-180:0:1 and 0.4);

\draw (0,1.5) node {{\color{red}$\bullet$}}; 

\draw[semithick] (5,1.1) arc (270:90:0.166 and 0.4);
\draw[dashed,color=gray] (5,1.1) arc (-90:90:0.166 and 0.4);

\draw (6,1.5) node {\LARGE{\color{red}$\bullet$}};

\draw[|-,semithick] (0,0) -- (4,0); \draw[|-|,semithick] (4,0) --
(6,0); \draw (0,-0.5) node {$x=0$}; \draw (4,-0.5) node
{$x=\frac{1}{3}$}; \draw (6,-0.5) node {$x=\frac{1}{2}$};
\end{tikzpicture}
\caption{The perturbative vacuum moduli space of the compactified
ISS model. The cones represent the classical Higgs branches. The
perturbative Coulomb branch shrinks near the points $x=0,1/3,1/2$
where charged fields become massless. The red neighborhoods are
where strong-coupling effects are expected to undermine
semi-classical techniques. \label{fig:pertVac}}
\end{figure}

We have now almost completely explained Figure~\ref{fig:pertVac}.
The one remaining feature, namely the pinches at $x=0,1/3,1/2$, are due
to the perturbative running of the gauge coupling, which sets the
radius of the dual photon; this radius shrinks near points where
charged fields become massless---see \cite{Strassler:2003qg} for a
clear exposition.\\

No further perturbative effects are expected to modify the vacuum
structure (because of the perturbative non-renormalization theorems in particular), so we now turn to the non-perturbative effects.

There are known semi-classical, non-perturbative effects on
$\mathbb{R}^3\times S^1$ due to BPS (anti-) monopoles
\cite{Gross:1980br} and KK (anti-) monopoles
\cite{Lee:1997vp,Kraan:1998pm}. These can be described in terms of a
low-energy \emph{monopole superfield}.

For SU($2$) gauge group, inside the classical Coulomb branch
parametrized by $x(=x_1)\in(0,1/2)$, and away from pinch-points
and the end-points of the interval, the (lowest component of the) monopole superfield takes
the form \cite{Affleck:1982as}
\begin{equation}
    Y\sim \exp\left(\frac{2x}{\hat{g}^2}\right),\label{eq:xToY}
\end{equation}
with $\hat{g}^2=g^2/8\pi^2$ and $g$ the gauge coupling. (We have
suppressed the dual photon for simplicity; restoring it would only give $Y$ a phase, and leave the modulus as in (\ref{eq:xToY}). See \cite{Aharony:2013dha}
for a recent, clear exposition of the relevant basics.) Note however that in $\mathcal{O}(g^2)$ neighborhoods of
pinch-points and the end-points of the interval, quantum-mechanical
effects are expected to modify the semi-classical relation (\ref{eq:xToY}) between $x$ and $Y$ \cite{Aharony:1997bx,Dorey:1998kq}.

An $n$-BPS-monopole effect \emph{may} contribute $1/Y^n$ to the low-energy
superpotential \cite{Affleck:1982as,Dorey:1998kq} on $\mathbb{R}^3\times
S^1$, and an $m$-KK-monopole effect \emph{may} contribute $(\eta Y)^m$,
where $\eta=\mathe^{-1/\hat{g}^2}$ is the 4d instanton factor. (Incidentally, a 4d
SU($2$) instanton is often thought of as a composite configuration
made out of a BPS and a KK monopole: $\frac{1}{Y}\times\eta
Y=\eta$.)

For such multi-monopoles to \emph{actually contribute} to the low-energy
superpotential though, it is necessary that they have precisely two fermionic
zero-modes \cite{Affleck:1982as}. The BPS and the KK monopole both have
precisely two gaugino zero-modes, so in absence of chiral multiplets
(i.e. in pure SU($2$) SYM) they both contribute to the
superpotential, which hence reads $1/Y+\eta Y$. In the ISS model we have to take into account the extra zero-modes arising from the
chiral multiplet in the spin-$3/2$ representation of the gauge
group. These zero-modes are counted by the following index theorem,
given by Poppitz and \"{U}nsal in \cite{Poppitz:2008hr} (see also
\cite{Nye:2000eg}):
\begin{equation}
    I^{\mathrm{BPS}}_{3/2}=10-I^{\mathrm{KK}}_{3/2}=-3\lfloor-3x\rfloor-\lfloor-x\rfloor+\lfloor x\rfloor+3\lfloor3x\rfloor.\label{eq:indThmISS}
\end{equation}
As indicated in \cite{Poppitz:2008hr} (below its Eq.~(3.10)), the
above index theorem implies that for $0<x<1/3$ there are 4 BPS
zero-modes---thus 6 KK zero-modes, while for $1/3<x<1/2$ there are
10 BPS
zero-modes---hence no KK zero-modes.

It appears like in their later paper \cite{Poppitz:2009kz}, Poppitz
and \"{U}nsal did not consider the range $1/3<x<1/2$. Since in this range
the chiral multiplet of the ISS model has no fermionic zero-modes on
a KK-monopole background, there should be an $\eta Y$ superpotential
generated on this part of the classical Coulomb branch of the model.
This superpotential yields a potential which increases with
increasing $Y$, and hence with increasing $x$; therefore its
qualitative behavior is compatible with the Rains function of the
ISS model in that range---c.f. Figure~\ref{fig:ISS}!

What about the range $0<x<1/3$? As the index theorem shows, both the
BPS and the KK monopole have too many zero-modes to generate a
superpotential there. At this point, a result of Poppitz and
\"{U}nsal can crucially guide us further \cite{Poppitz:2009kz}: they
found that in the range $0<x<1/3$ the monopole superfield $Y$ has U($1$)$_R$ charge $-2/5$.
Thus the natural candidate for a superpotential term would be
$1/Y^5$, signaling a five-BPS-monopole effect. At first glance it
seems like such a term is forbidden because a five-BPS-monopole
background would have $5\times4=20$ chiral-multiplet zero-modes and
$5\times2=10$ gaugino zero-modes---far too many. However, we now
argue that the extra zero-modes are not protected by any symmetry,
so are expected to be lifted by quantum effects! In particular,
because the U($1$)$_R$ charge of the chiral-multiplet fermions is
$3/5-1=-2/5$ and that of the gauginos is $+1$, the 20
chiral-multiplet zero-modes can combine with 8 of the gauginos and
be lifted without violating U($1$)$_R$; the corresponding vertex
$\psi^{20}\lambda^8$ has U($1$)$_R$-charge
$20\times(-2/5)+8\times(+1)=0$, so is not prohibited by any
symmetry, and should therefore ``naturally''---in the technical
sense---be generated. We hence propose that \emph{the low-energy K\"{a}hler potential arising from the supersymmetric
sigma model on the moduli space of a five-BPS-monopole configuration
has such a vertex}, or actually the more fundamental
$\psi^{10}\lambda^4$ vertex which would also do the job of soaking up the extra zero-modes (c.f. section~3 of \cite{Gauntlett:1995fu} pertinent to an analogous four-dimensional situation). The vertex presumably couples the fermion zero-modes to some power of the curvature tensor of the index bundle \cite{Manton:1993aa} over the five-monopole moduli space (c.f. the last term in Eq.~(3.8) of \cite{Gauntlett:1995fu}).

Although our proposed mechanism for lifting of the extra zero-modes
might appear contrived, in fact several effects of similar nature
have already been observed both in three and four dimensions, with
various amounts of supersymmetry---see e.g.
\cite{Gauntlett:1995fu,Dorey:1997ij,Dorey:1998kq,Dorey:1999sj}. For the sceptic readers
we point out that the non-perturbative two-instanton vertex (c.f.
\cite{Poppitz:2009kz})
\begin{equation}
\mathe^{-2S_{\mathrm{inst}}}\psi^{20}\lambda^8,\label{eq:twoInstVertex}
\end{equation}
also does the job of lifting the extra zero-modes via K\"{a}hler corrections; absent the perturbative vertex of the previous paragraph, the vertex (\ref{eq:twoInstVertex}) would lead to a five-monopole effect suppressed by an $\eta^2=\mathe^{-2S_{\mathrm{inst}}}$
factor, but that suppression would not significantly modify our discussion
below.\footnote{However, the two scenarios imply different dynamics for the dimensionally reduced (3d) ISS model. If the perturbative vertex of the previous paragraph exists, its dimensional reduction would serve---analogously to its 4d parent on $\mathbb{R}^3\times S^1$---to lift the extra fermionic zero-modes of the five-instanton configuration in the 3d ISS model. (Note that the 3d five-instanton configuration is the dimensional reduction of the 4d five-BPS-monopole configuration. Moreover, its fermionic zero-modes are counted by the Callias index theorem \cite{Callias:1977kg}, which agrees with (\ref{eq:indThmISS}) near $x=0$, hence yielding four $\psi$ zero-modes per instanton and thus 20 in total in the present case). Therefore a $1/Y^5$ runaway superpotential will be generated on the Coulomb branch of the 3d model. If, on the other hand, the perturbative vertex is absent and only the non-perturbative vertex (\ref{eq:twoInstVertex}) originating from 4d instantons exists, since in the reduction limit ($r_{S^1}\to0$ with $g_3$ fixed, so that $\eta=\mathe^{-1/\hat{g}^2}=\mathe^{-1/r_{S^1}\hat{g}_3^2}\to0$) the vertex (\ref{eq:twoInstVertex}) does not yield a 3d vertex to lift the extra zero-modes in the dimensionally reduced model, there would not be any superpotential generated, and the 3d model would have an unlifted quantum Coulomb branch. Since our naturalness argument in the previous paragraph leads us to believe the perturbative vertex exists, we consider the runaway scenario more likely for the 3d ISS model.}

We conclude that a quintic monopole-superpotential of the form
$1/Y^5$ is indeed generated over the range $0<x<1/3$ of the
classical Coulomb branch. This potential decreases with increasing
$x$; therefore its qualitative behavior is compatible with the Rains
function of Figure~\ref{fig:ISS} in that range! (Some similar higher-power monopole superpotentials have been recently used in \cite{Benini:2017dud,Amariti:2018gdc,Amariti:2019rhc}---though only up to quadratic degree.)\\

It is useful---for holomorphy arguments in particular---to employ
different Coulomb-branch operators on different sides of the
pinch-points. Thus for $x>1/3$ instead of $Y$ we may use $Y'_+\sim
\exp\left(\frac{2x'}{\hat{g}^2}\right)$, with $x':=x-1/3$. At the quantum-mechanical level we expect $Y'_+$ to have the nice property that
$Y'_+\to0$ as $x'\to0^+$ \cite{Aharony:1997bx}. On the other hand, on a
neighborhood to the left of $x=1/3$, the Coulomb-branch operator
$Y'_-\sim \exp\left(-\frac{2x'}{\hat{g}^2}\right)$, proportional to
$1/Y$, becomes useful as we expect that quantum-mechanically
$Y'_-\to 0$ as $x'\to0^-$. Therefore we can summarize our findings
in this section by writing the low-energy superpotential on the
classical Coulomb branch of the ISS model on $\mathbb{R}^3\times
S^1$ as
\begin{equation}
W\propto\begin{cases}Y'^5_- & 0<x<1/3,\\
Y'_+ & 1/3<x<1/2.
\end{cases}\label{eq:semiClassicW}
\end{equation}
This superpotential---previously thought to be zero---is the main result of the present paper.

Over the range $0<x<1/3$ we argued that a vertex of the form
$\psi^{20}\lambda^8$ (or two vertices of the form
$\psi^{10}\lambda^4$) should \emph{naturally} lift the extra
fermionic zero-modes of a five-BPS-monopole background. For
$1/3<x<1/2$ we simply used an earlier result of Poppitz and
\"{U}nsal \cite{Poppitz:2008hr} that the KK-monopole background does
not have extra fermionic zero-modes. In particular, the semi-classical vacuum structure implied by the
low-energy superpotential (\ref{eq:semiClassicW}) is compatible with
that suggested by the Rains function (\ref{eq:LhISS}) shown in
Figure~\ref{fig:ISS}.

Note that the superpotential (\ref{eq:semiClassicW}) is useful only outside a small neighborhood of $x=1/3$ wherein new light states associated with $q_{\pm3}$ appear. However, the superpotential seems to drive the theory to a vacuum in that small neighborhood. The nature of this vacuum is not entirely clear to us, and we will only present some preliminary comments on it in the next section.\\

It might be possible that a perspective from Seiberg-Witten theory
\cite{Seiberg:1996nz} (see also \cite{Aharony:1997bx}) can shed further
light on the above picture once the ISS model is embedded in the
$\mathcal{N}=2$ $I^\ast_1$ theory of
\cite{Argyres:2010py,Argyres:2015ffa}. This approach is currently under
investigation.

\section{The low-energy phase}\label{sec:3dIR}

\noindent In this section we make some rudimentary remarks concerning the low-energy phase, arguing in particular that the theory has a supersymmetric, gapless vacuum. A more careful, conclusive study is left for future work.

First, following the conventional wisdom we assume that the red
neighborhoods of the $x=0,1/2$ points, although not really amenable
to semi-classical analysis, presumably contain no acceptable vacua
because in their semi-classical vicinity the low-energy effective
potential is repulsive---c.f. the case of 3d pure $\mathcal{N}=2$
SYM for instance \cite{Aharony:1997bx}. Hence below we discuss only
the vacuum structure near $x=1/3$.

The SU($2$) gauge group is broken down to U($1$) of course.
Moreover, of the four components of the chiral multiplet
$Q_{I=3/2}$, having U($1$) charges $-3,-1,+1,+3$, the two with
charges $-1,+1$ do not yield massless 3d descendants near $x=1/3$. In the
resulting low-energy theory we are thus left with the field content of 3d
$\mathcal{N}=2$ SQED with chiral multiplets $q_{-3},q_{+3}$ of
charge $-3,+3$ coming from the KK modes $\pm1$ of their 4d parents.
This is in fact compatible with the \emph{subleading} Cardy-like
asymptotics of the Romelsberger index of the model as described in
appendix~\ref{appendix:FI}.

It is important to note that the field content discussed in the previous paragraph, and the effective SQED predicated on it, provide a useful description only up to a cut-off scale (or ``threshold'')
\begin{equation}
\Lambda^{\mathrm{3d\ SQED}}_{\mathrm{uv}}=\frac{\epsilon}{r_{S^1}},\label{eq:epsilonCut}
\end{equation}
with some $\epsilon\ll 1$. That is to say we consider this SQED as an effective description of only an $\mathcal{O}(\epsilon)$ neighborhood\footnote{The use of this epsilon cut-off is motivated by a similar epsilon cut-off arising in the asymptotic analysis of the index \cite{Ardehali:2015bla}.} of the pinch at $x'(:=x-1/3)=0$. Outside this $\mathcal{O}(\epsilon)$ neighborhood---and away from the red regions---the U(1) gauge theory of the previous section (with the superpotential deformation (\ref{eq:semiClassicW})) provides the appropriate effective description.

In this SQED theory near $x=1/3$, there are no FI or Chern-Simons
terms induced through one-loop effects, because the massive fields
come in pairs of opposite charge under the low-energy U($1$) gauge
group. Non-perturbative effects presumably cannot induce a properly
quantized Chern-Simons level either, but since FI parameters need not be quantized we leave open the
possibility of an instanton-induced FI parameter $\zeta$. (As
we will discuss in appendix~\ref{appendix:FI}, the subleading
$\beta\to0$ asymptotics of the $S^3\times S^1$ partition function
signals an induced `imaginary FI parameter', which we will explain in that appendix from the
$r_{S^1}\to0$ perspective via ``high-temperature'' perturbative
effects; it is the possibility of an induced FI parameter from the
$\mathbb{R}^3\times S^1$ perspective via ``low-energy''
non-perturbative effects that we are leaving open.)

We expect that the near-pinch SQED inherits a superpotential of the form (\ref{eq:semiClassicW}), which would lift its Coulomb branch. In fact one might also expect an extra term of the form $M{Y'}_-^{1/3}{Y'}_+^{1/3}$ to be generated dynamically in the theory, since it has the right R-charge: recalling that $q_{\pm3}$ have R-charge $3/5$, the R-charge of $M\sim q_{+3}q_{-3}$ becomes $6/5$, while those of $Y'_{-,+}$ follow from (\ref{eq:semiClassicW}) to be $2/5,2$, therefore $M{Y'}_-^{1/3}{Y'}_+^{1/3}$ has R-charge $6/5+2/15+2/3=2$. Hence we propose 
\begin{equation}
W_{\mathrm{SQED}}\sim M{Y'}_-^{1/3}{Y'}_+^{1/3} +Y'^5_- + Y'_+,\label{eq:WirSQED}
\end{equation}
wherein $Y'_{+},Y'_{-},M$ ought to be considered independent operators. We now proceed to present independent arguments supporting a slightly more refined version of (\ref{eq:WirSQED}), given in (\ref{eq:XYZsup}).\footnote{We are indebted to O.~Aharony for helpful correspondences regarding the subject of the present section.}

A point we need to establish here is that the near-pinch SQED not only has extra massless fields compared to the effective U(1) theory outside the $\mathcal{O}(\epsilon)$ neighborhood, it also has different monopole operators. This is seen from the following discussion.

Recall that the chiral multiplet $Q_{I=3/2}$ yields states of charge $-3,-1,+1,+3$ on the Coulomb branch. However, because the states of charge $\pm1$ have masses $\ge1/3r_{S^1}$, \emph{they lie far beyond the SQED cut-off $\epsilon/r_{S^1}$}. Therefore we need not worry about their quantization in the near-pinch SQED. It is only the electric sources with charge $\pm3$ that have to satisfy a Dirac quantization condition with respect to the monopoles, implying that the minimal allowed magnetic charge is $\pm1/3$. The minimal-flux monopole operators of the near-pinch SQED hence read (c.f. \cite{Aharony:2013dha})
\begin{equation}
    V_{\pm}\sim \exp{\left[\pm\frac{2\tilde{x}'}{3\hat{g}_3^2}\pm\frac{2ia}{3}\right]},\label{eq:Vpm}
\end{equation}
with $a$ the dual photon (which we restore in this section), and $\tilde{x}'$ the 3d SQED Coulomb-branch parameter (to be identified with $x'/r_{S^1}$ near the threshold).

As a sanity check, we note that the effective U(1) theory valid outside the $\mathcal{O}(\epsilon)$ neighborhood of the pinch has cut-off scale
\begin{equation}
\Lambda^{\mathrm{3d\ U(1)}}_{\mathrm{KK}}=\frac{1}{r_{S^1}}.\label{eq:U(1)cutoff}
\end{equation}
The charge-$\pm1$ states should therefore \emph{be included} in the U(1) theory away from the pinch, since their masses (the lowest of which is $1/3r_{S^1}$) do not lie far beyond the U(1) theory cut-off (\ref{eq:U(1)cutoff}). So indeed the minimal magnetic charge is $\pm1$, and the minimal-flux operators for generic distances $|x'|=\mathcal{O}(\epsilon^0)$ away from the pinch are:
\begin{equation}
    Y'_{\pm}\sim \exp{\left[\pm\frac{2x'}{\hat{g}^2}\pm2ia\right]},\label{eq:Yprime}
\end{equation}
 just as claimed in the previous section, but now with $a$ reinstated.

Recalling that $\hat{g}^2=r_{S^1}\hat{g}_3^2$, the forms of (\ref{eq:Yprime}) and (\ref{eq:Vpm}) suggest that at the threshold $x'=\pm\epsilon\leftrightarrow \tilde{x}'=\frac{\pm\epsilon}{r_{S^1}}$ we identify
\begin{equation}
    Y'_{\pm}\sim V_{\pm}^3.
\end{equation}
(If the SQED had charges $\pm n$, the exponent of the RHS would be $n$.)

Now, matching at the threshold with the superpotential (\ref{eq:semiClassicW}) of the U(1) theory away from the pinch suggests taking
\begin{equation}
    W_{\mathrm{uv\ SQED}}\sim V_-^{15}+V_+^3.
\end{equation}
However, we expect from the SQED/XYZ duality \cite{Aharony:1997bx} that in the infrared our (deformed) SQED is alternatively described via a (deformed) XYZ model. We propose that the natural identification would be between $X,Y$ and the \emph{minimal-flux operators} $V_\pm$, as well as between $Z$ and $M$ per usual \cite{Aharony:1997bx}. The (deformed) XYZ superpotential would then read
\begin{equation}
    W_{XYZ}=XYZ+X^{15}+Y^3.\label{eq:XYZsup}
\end{equation}
Interpreting the $XYZ$ term in (\ref{eq:XYZsup}) as a dynamically generated $MV_- V_+$ term in the SQED picture would then justify (\ref{eq:WirSQED}). The advantage of (\ref{eq:XYZsup}) over (\ref{eq:WirSQED}) is that it involves the appropriate monopole operators $X,Y$ (or altarnatively $V_\pm$) of the near-pinch SQED, rather than the monopole operators $Y'_{\pm}$ of the U(1) theory valid away from the pinch.

We now observe that the superpotential (\ref{eq:XYZsup}) (or alternatively (\ref{eq:WirSQED}) when written in terms of $V_{\pm},M$) is stationary at $X=Y=0$ (alternatively at $V_\pm=0$). Therefore the low-energy phase appears to be \emph{supersymmetric}.

For nonzero
$\zeta$, the superpotential (\ref{eq:XYZsup}) moreover implies that our SQED would have a
Higgs branch parametrized by vevs of the two chiral-multiplet
scalars subject to $|q_{+3}|^2-|q_{-3}|^2=\zeta/6\pi$ (c.f.
section~2 of \cite{Intriligator:2013lca}); if $\zeta=0$ on the
other hand, the $M=0$ point becomes a viable vacuum as well. Either way, the low-energy phase appears to be \emph{gapless}.

A discussion of gapped phases of somewhat similar models can be found in \cite{Strassler:1999hy}.

\section{Discussion: what are exact results good
for?}\label{sec:discussion}

\noindent It is often said that exact results are valuable because
they \emph{reach beyond semi-classical techniques}. One of the main purposes
of the present article has been to advocate the program of using exact results rather to \emph{shed further light on semi-classical regimes}: exact results might
inform us of subtle effects that may have escaped our
earlier semi-classical studies. We demonstrated this program in action by leveraging the exact Romelsberger index of the ISS model to uncover subtle non-perturbative effects overlooked in the earlier semi-classical analysis of its vacua on $\mathbb{R}^3\times S^1$.

With this vision in mind, one can
begin comparing the semi-classical results on arbitrary
$\mathcal{N}=1$ gauge theories with a U($1$)$_R$ symmetry with the
behavior expected from the index perspective. As discussed in section~\ref{sec:Rains}, in many
cases the two perspectives are compatible, at least as far as the
dimension of the implied quantum Coulomb branch is concerned; see also
subsection~5.4 of \cite{Ardehali:2015bla}. There are, however, certain potential obstructions to compatibility of the two perspectives in general, which we now outline.

A first potential obstruction is that the limit $r_{S^3}\to\infty$ may not correctly capture the dynamics ``at $r_{S^3}=\infty$''. This can arise because certain non-generic properties influencing the dynamics appear only at $r_{S^3}=\infty$. Enhanced supersymmetry is one example of such non-generic properties. A typical $\mathcal{N}=2$ SCFT has a non-vanishing Rains function (take the $\mathbb{Z}_2$ orbifold theory discussed in \cite{Ardehali:2015bla} as a specific instance), while its classical Coulomb branch on $\mathbb{R}^3\times S^1$ is protected by the enhanced supersymmetry. This incompatibility can be traced back to the fact that at arbitrary large but finite $r_{S^3}$ the background breaks the $\mathcal{N}=2$ SUSY protecting the Coulomb branch, while at $r_{S^3}=\infty$ the extended supersymmetry is recovered. (The $\mathcal{N}=4$ theory is somewhat special in this respect: its Rains function vanishes, just as expected from the enhanced SUSY protecting its classical Coulomb branch, so the obstruction does not ruin the connection in this case \cite{Ardehali:2015bla}. It essentially has so much supersymmetry that even at finite $r_{S^3}$ its Coulomb branch is protected.)

A second potential obstruction is that the index might not contain all the information relevant to the dynamics of the theory on $\mathbb{R}^3\times S^1$. For instance, the index is insensitive to various ``global'' subtleties (such as the spectrum of the line operators \cite{Aharony:2013hda}), so might not capture the dynamics on $\mathbb{R}^3\times S^1$ when such global issues become crucial in determination of the vacua.

At present, there are no general criteria for telling in advance if one of the obstructions mentioned above undermines the connection between the Rains function and the $\mathbb{R}^3\times S^1$ vacua. The available strategy is to focus on cases that are expected to be free of various ``accidents'' (such as SUSY enhancement) or global subtleties, take the Rains function seriously as a guide to the $\mathbb{R}^3\times S^1$ dynamics, and see where it leads us. This is the path that we followed with the ISS model in the present paper.

When the above obstructions are absent (as in the ISS model or various SQCD theories) and the minima of the Rains function do correspond to the $\mathbb{R}^3\times S^1$ vacua, the remaining open question is the precise connection between the Rains function (which is piecewise linear) and the semi-classical low-energy
superpotential on $\mathbb{R}^3\times S^1$ (which is piecewise exponential). So far a satisfactory semi-classical understanding of the
Rains function exists only in the ``direct channel'' where the
$\beta\to0$ limit is interpreted as shrinking the $S^1$
\cite{DiPietro:2016ond} rather than decompactifying the $S^3$; there,
it originates from \emph{perturbative effects in high-temperature
effective field theory}. Here, as in \cite{Ardehali:2015bla}, we have
suggested that the Rains function might also encode
\emph{non-perturbative effects} in the ``crossed channel'', i.e. on
$\mathbb{R}^3\times S^1$. Such a remarkable connection between
perturbative effects in the direct channel and non-perturbative
effects in the crossed channel would be reminiscent of strong-weak
duality, and a more systematic investigation of it would be quite worthwhile.

An alternative ``crossed-channel'' perspective, due to Shaghoulian
\cite{Shaghoulian:2016gol}, is given on the Rains function in
appendix~\ref{appendix:Shaghoulian}. It involves a rather
heuristic argument relating $S_{r_{S^1}\to0}^1\times S^3$ to
$S^1\times S^3/\mathbb{Z}_{p\to\infty}$, and does not appear to bear
implications for our main discussions in this paper. We
nevertheless find it an exciting complementary angle worth
further exploration.

Finally, what (if any) implications our results might have for the gauge
dynamics of the ISS model on $\mathbb{R}^4$ remains to be
understood.

\acknowledgments

We would like to thank E.~Poppitz and M.~\"{U}nsal for their helpful
feedbacks and encouraging remarks at various stages of this project.
The analysis in appendix~\ref{appendix:FI} in particular originated
from conversations with E.~Poppitz. We are indebted to O.~Aharony and S.~Razamat for enlightening correspondences and suggestions regarding 3d SQED, and to A.~Amariti, L.~Di~Pietro, and G.~Festuccia for constructive feedback on a draft of this work. We are also thankful to
F.~Benini, A.~Bourget, G.~Festuccia, Z.~Komargodski, J.~Minahan,
U.~Naseer, S.~Pasquetti, and E.~Shaghoulian for related discussions, as well as to two anonymous JHEP referees whose feedback on an earlier draft of this manuscript was essential to its subsequent improvement.
This work was supported by the Knut and Alice Wallenberg Foundation
under grant Dnr KAW 2015.0083. The work of L.C. was supported in part
by Vetenskapsr\r{a}det under grant \#2014-5517, by the STINT grant,
and by the grant ``Geometry and Physics'' from the Knut and Alice Wallenberg foundation.

\appendix

\section{Physics of the subleading asymptotics}\label{appendix:FI}

\noindent The exact expression for the index of the ISS model reads
\begin{equation}
\mathcal{I}_{\mathrm{ISS}}(\beta)=\frac{(q;q)^2_\infty}{2}\int_{-1/2}^{1/2} \mathrm{d}x\frac{\Gamma_e(q^{3/5}z^{-3})\Gamma_e(q^{3/5}z^{-1})\Gamma_e(q^{3/5}z^{+1})\Gamma_e(q^{3/5}z^{+3})}{\Gamma_e(z^{-2})\Gamma_e(z^{+2})}.\label{eq:issIndexEHI}
\end{equation}
Here $z=\mathe^{2\pi i x}$, $q=\mathe^{-\beta}$, $(x;q)_\infty:=\prod_{k=0}^\infty (1-x q^k)$ is the $q$-Pochhammer symbol, and $\Gamma_e$ is the elliptic gamma function \cite{Ruijsenaars:1997}. The expression in (\ref{eq:issIndexEHI}) is the specialization to $p=q$ of the corresponding two-variable elliptic hypergeometric integral written down first by Vartanov \cite{Vartanov:2010xj}.  Because of the power 3/5 of $q$ in the arguments of the elliptic gammas in the numerator, we expect $\mathcal{I}_{\mathrm{ISS}}$ to be a meromorphic function of $q$ on the \emph{quintuple branched cover} of the punctured unit open disk---c.f. the appendix of \cite{Rains:2005}. As in the rest of this work though, we continue restricting to $\beta\in\mathbb{R}^{>0}$. A simple argument then establishes that $\mathcal{I}_{\mathrm{ISS}}$ is a continuous real function \cite{ArabiArdehali:2017fsp}.

In \cite{Ardehali:2015bla} it was found that (c.f.
Eq.~(3.80) there)
\begin{equation}
\mathcal{I}_{\mathrm{ISS}}(\beta)\simeq
\mathe^{\frac{16\pi^2}{3\beta}\frac{1}{80}}\cdot
Y_{S^3}^{\mathrm{ISS}}\cdot \mathe^{\beta
E_{\mathrm{susy}}^{\mathrm{ISS}}}\qquad (\text{as
}\beta\to0),\label{eq:issIndexAsy}
\end{equation}
where $E_{\mathrm{susy}}^{\mathrm{ISS}}=511/12,$ and
\begin{equation}
Y^{\mathrm{ISS}}_{S^3}=\int_{-\infty}^{\infty}\mathrm{d}x' \mathe^{-2\pi
i(-\frac{4}{5}i)x'}\times\Gamma_h(3x'+3i/5)\Gamma_h(-3x'+3i/5),\label{eq:Yiss}
\end{equation}
with $\Gamma_h$ the hyperbolic gamma function. (One can of course numerically evaluate
$Y^{\mathrm{ISS}}_{S^3}\approx0.423$ \cite{Ardehali:2016kza}, but
here we are interested in the physical interpretation of the integral
representation of $Y^{\mathrm{ISS}}_{S^3}$.) The precise meaning of the symbol $\simeq$ in (\ref{eq:issIndexAsy})
is that after taking logarithms of the two sides we get an equality
to all orders in $\beta$.

As alluded to in the main text, Di~Pietro and Honda
\cite{DiPietro:2016ond} have given a semi-classical explanation of the
leading asymptotics---the first factor on the RHS of
(\ref{eq:issIndexAsy})---in the ``direct channel'' where the Cardy
limit is interpreted as shrinking the circle of $S^3\times S^1$.
This was done through high-temperature effective field theory,
building on Di~Pietro and Komargodski's work \cite{DiPietro:2014bca}.

A high-temperature effective field theory explanation for the
$E_{\mathrm{susy}}$ piece---i.e. the third factor on the RHS of
(\ref{eq:issIndexAsy})---follows from the more recent work
\cite{Closset:2019ucb}.

The purpose of this appendix is to provide a similar
``direct-channel'' explanation for the $Y^{\mathrm{ISS}}_{S^3}$
piece. Of course $Y^{\mathrm{ISS}}_{S^3}$ looks like the $S^3$
partition function of the SQED arising after the SU($2$)$\to$U($1$)
Higgsing driven by the Rains function---compare for instance with
the expressions in section~5 of \cite{Aharony:2013dha}. The only part
of it deserving further explanation is the factor $\mathe^{-2\pi
i(-\frac{4}{5}i)x'}$ in the integrand, which as noted in
section~5 of \cite{Ardehali:2015bla} appears to signal an induced FI
parameter. The semi-classical explanation of this `imaginary FI parameter'\footnote{Note that the usual FI parameters correspond to \emph{real} vev for the sigma field of the background vector multiplet, so our references to an \emph{imaginary} FI parameter might be considered to involve a bit of an abuse of terminology. Whether this distinction suggests subtleties in a proper crossed-channel interpretation of this `FI parameter' is not clear to us. We thank L.~Di~Pietro for an instructive correspondence on this point.} $\zeta=-\frac{4}{5}i$ is the subject of the rest of this appendix. (See section~3 of
\cite{Amariti:2014lla} for a related discussion).

On the Coulomb branch of an SU($2$) theory---in the Weyl chamber
with $x>0$ to be concrete---massive fermions can be integrated out.
Upon integrating out each of these massive fermions, a one-loop
mixed gauge-U($1$)$_R$ Chern-Simons term is generated with
coefficient
\begin{equation}
\frac{1}{2}R\cdot n \cdot\mathrm{sign}(nx),
\end{equation}
where $R$ is the U($1$)$_R$ charge, and $n$ is the U(1) gauge charge
of the fermion. More precisely, each such fermion is accompanied by
a tower of KK modes which modify the above coefficient to
\begin{equation}
\sum_{m\in\mathbb{Z}}\frac{1}{2}R\cdot n
\cdot\mathrm{sign}(nx+m)=R\cdot n\cdot(1/2-\{nx\})\qquad  \text{for
}nx\notin\mathbb{Z}.\label{eq:mixedCS}
\end{equation}
Here we have used the regularizations
$\sum_{m\in\mathbb{Z}}\mathrm{sign}(m+T)=1-2\{T\}$ valid for
$T\notin\mathbb{Z}$. Note that in the case $nx\in\mathbb{Z}$ where
there is a massless descendant, the regularization
$\sum_{m\in\mathbb{Z}}\mathrm{sign}(m+T)=0$ valid for
$T\in\mathbb{Z}$ guarantees that the massive modes in its KK tower
do not contribute.

The mixed Chern-Simons coefficient (\ref{eq:mixedCS}) appears as an
FI term in the (high-temperature) effective Lagrangian:
\begin{equation}
\zeta_{R,n}=R\cdot n\cdot(1/2-\{nx\})\omega.\label{eq:indFI}
\end{equation}
Here $\omega$ is determined by the background fields that couple the
U($1$)$_R$ current multiplet (c.f. Section~2 and Appendix~A of \cite{DiPietro:2016ond}). For $S^3$ it is $\omega=i$. (Setting
$\omega=i(b+b^{-1})/2$ instead, generalizes the story to squashed
$S^3$ with squashing parameter $b$.)

The expression (\ref{eq:indFI}) in fact matches the FI coefficient
arising from the Cardy limit of the index of a single chiral
multiplet---c.f. the estimate (3.53) in \cite{Ardehali:2015bla}. To see
specifically how the FI parameter $\zeta=-\frac{4}{5}i$ in \eqref{eq:Yiss} is
generated, we note that (\ref{eq:indFI}) in this case implies
\begin{equation}
\begin{split}
\sum_{\mathrm{fer.\ with\ }nx\notin\mathbb{Z}}R\cdot
n\cdot(1/2-\{n\frac{1}{3}\})i&=1\cdot 2\cdot (1/2-\{\frac{2}{3}\})i+1\cdot
(-2)\cdot(1/2-\{-\frac{2}{3}\})i\\
&\ \ \ +(-\frac{2}{5})\cdot 1\cdot(1/2-\{\frac{1}{3}\})i+(-\frac{2}{5})\cdot
(-1)\cdot(1/2-\{-\frac{1}{3}\})i\\
&=-\frac{4}{5}i,\label{eq:indFIiss}
\end{split}
\end{equation}
with the first line on the RHS coming from the gauginos which have
gauge charges $n=\pm2$ and R-charge $R=1$, and the second line
coming from the two chiral-multiplet fermions with $n=\pm1$ and
$R=-2/5$. The other two chiral-multiplet fermions with $n=\pm3$ have
$nx=\pm3\cdot\frac{1}{3}\in\mathbb{Z}$, and hence (according to the remarks below (\ref{eq:mixedCS})) do not contribute
to the FI parameter.

\section{Rains function from Shaghoulian's perspective}\label{appendix:Shaghoulian}

\noindent Shaghoulian has conjectured that the \emph{leading}
asymptotics of the supersymmetric partition functions on
$S_{r_{S^1}\to0}^1\times S^3$ and $S^1\times
S^3/\mathbb{Z}_{p\to\infty}$ are equal upon the ``modular''
identification
\begin{equation}
\frac{\tilde{r}_{S^3}/p}{\tilde{r}_{S^1}}=\frac{r_{S^1}}{r_{S^3}},\label{eq:modId}
\end{equation}
with the tilded parameters those of the latter space
\cite{Shaghoulian:2016gol}. The (rather heuristic) reasoning behind the
conjecture is essentially as follows: both $S_{r_{S^1}}^1\times S^3$
and $S^1\times S^3/\mathbb{Z}_{p}$ are torus-bundles over $S^2$; the
limit $r_{S^1}\to0$ shrinks one cycle of the torus, while
$p\to\infty$ shrinks the other, and modularity relates the two
limits; effects of non-trivial fibration  undermine the asymptotic
equality of the partition functions at subleading orders, but
(conjecturally) not at the leading order.

A Cardy-like \cite{Cardy:1986ie} argument then gives the leading
small-$\beta$ asymptotics of the $S^1\times S^3$ partition function
in terms of the supersymmetric Casimir energy $E_{\mathrm{susy},p}$
on $S^1\times S^3/\mathbb{Z}_{p}$. Shaghoulian produced the
$\mathe^{-\mathcal{E}_0^{\mathrm{DK}}}$ factor in (\ref{eq:indexAsy}) by
appealing to the supersymmetric Casimir energy in the zero-holonomy
sector of $S^1\times S^3/\mathbb{Z}_{p}$ \cite{Shaghoulian:2016gol}.
Below we show how the $\mathe^{-\frac{4\pi^2}{\beta}L_h}$ piece in
(\ref{eq:indexAsy}), and hence the Rains function, arise when the
dependence of $E_{\mathrm{susy},p}$ on the ``spatial'' holonomies
around the non-trivial cycle of $S^3/\mathbb{Z}_{p}$ is taken into
account.\footnote{This incidentally demystifies an observation made in
footnote~4 of \cite{Ardehali:2015bla}, which is actually incorrect as
it stands, since it refers to the $p\to1$ limit rather than the
correct $p\to\infty$ limit.}

The starting point is Eq.~(5.6) of Martelli and Sparks
\cite{Martelli:2015kuk} for the supersymmetric Casimir energy on
$S^1\times S^3/\mathbb{Z}_p$. We are interested in the $p\to\infty$
limit, and in the round $S^3$ case corresponding to $b_1=b_2=1$ in \cite{Martelli:2015kuk}. We
begin by considering a chiral multiplet $\chi$ with $R$-charge
$r_\chi$. Below we denote $\boldsymbol{m}/p$ in that work by
$\tilde{\boldsymbol{x}}$, and hence $\nu/p$ in that work by
$\{\rho^\chi\cdot\tilde{\boldsymbol{x}}\}$. Eq.~(5.6) of
\cite{Martelli:2015kuk} then immediately gives (from its linear term in
$u=r_\chi-1$)
\begin{equation}
\frac{E^\chi_{\mathrm{susy},p}(\tilde{\boldsymbol{x}})}{p}\overset{p\to\infty}{\longrightarrow}
\frac{r_\chi-1}{12}+\frac{1}{2}(1-r_{\chi})\vartheta(\rho^\chi\cdot\tilde{\boldsymbol{x}}).
\end{equation}

By summing over all the chiral multiplets and also
including the vector multiplets the above relation becomes
\begin{equation}
\frac{E_{\mathrm{susy},p}(\tilde{\boldsymbol{x}})}{p}\overset{p\to\infty}{\longrightarrow}
\frac{\mathrm{Tr}R}{12}+L_h(\tilde{\boldsymbol{x}}).
\end{equation}

The ``vacuum'' energy is of course obtained by minimizing
$E_{\mathrm{susy},p}(\tilde{\boldsymbol{x}})$ over the moduli space
of the ``spatial'' holonomies $\tilde{\boldsymbol{x}}$. For this
minimized value, $E_{\mathrm{susy},p}^{\mathrm{min}}$, we get
\begin{equation}
\frac{E^{\mathrm{min}}_{\mathrm{susy},p}}{p}\overset{p\to\infty}{\longrightarrow}
\frac{\mathrm{Tr}R}{12}+L_{h}^{\mathrm{min}}.
\end{equation}

Since from (\ref{eq:modId}) we have $\tilde{\beta}=4\pi^2/p\beta$,
Shaghoulian's conjecture now implies:
\begin{equation}
\log Z(S^1\times
S^3)\overset{\beta\to0}{\longrightarrow}\lim_{p\to\infty}-\frac{4\pi^2}{p\beta
}E^{\mathrm{min}}_{\mathrm{susy},p}=-\frac{\pi^2}{3\beta}(\mathrm{Tr}R+12
L_{h}^{\mathrm{min}}),\label{eq:correctAsy}
\end{equation}
in complete accord with (\ref{eq:indexAsy}).

When $L_{h}^{\mathrm{min}}=0$ we of course recover the
Di~Pietro-Komargodski formula; this happens when
$E_{\mathrm{susy},p}(\tilde{\boldsymbol{x}})$ is minimized in the
zero-holonomy sector, since $L_h(0)=0$. But in the case of the ISS
model, \emph{Higgs vacua} with $\tilde{\boldsymbol{x}}\neq0$
minimize $E_{\mathrm{susy},p}(\tilde{\boldsymbol{x}})$, thereby
modifying the Di~Pietro-Komargodski asymptotics.

\end{document}